
\input harvmac
\def\b{\bar}
\def\t{\tilde}
\def\h{N_h}
\def\g{N_g}

\Title{\vbox{\baselineskip12pt\hbox{CERN-TH-6433/92}\hbox{NEIP92-002}}}
{\vbox{\centerline{Supersymmetry Protects the Primordial}
   \vskip2pt\centerline{Baryon Asymmetry }}}

\centerline{Luis E. Ib\'a\~nez\footnote{$^1$}
{Address after 1 October 1992, Departamento de  F\'isica
Te\'orica C-XI, Universidad Aut\'onoma de Madrid, Cantoblanco,
28049, Madrid, Spain;
email: ibanez@cernvm}}
\centerline{CERN, 1211 Geneva 23, Switzerland}
\bigskip\centerline{and}
\bigskip\centerline{Fernando Quevedo\footnote{$^2$}{
Supported by the Swiss National Science
                                Foundation, email: quevedo@iph.unine.ch}}
\centerline{Institut de Physique, Universit\'e de Neuch\^atel}
\centerline{Rue A.-L. Breguet, 1 CH-2000, Switzerland}

 \vskip .2in
 \noindent
\vbox{\baselineskip12pt
It has been argued that any primordial $B+L$ asymmetry existing
at very high temperatures  can be
subsequently erased by anomalous electroweak effects. We argue
that this is not  necessarily the case in the supersymmetric standard model
because, apart from $B$ and/or $L$, there are, above
 a certain temperature $T_{SS}$, two other anomalous $U(1)$ currents.
  As a consequence,
anomalous
electroweak effects are only able to partially transform a $B+L$ excess
into a generation of primordial sparticle (e.g. gaugino) density.
This relaxes recent bounds on $B,L$-violating non-renormalizable
couplings by several orders of magnitude. In particular, dimension-5
  couplings inducing neutrino masses may be 4 orders of magnitude
larger than in the non-supersymmetric case, allowing for neutrino
masses  $m_{\nu }\leq 10    $ eV. These values are consistent
with a MSW+see-saw explanation of the solar-neutrino data and also
with  possible  $\nu _{\mu }\leftrightarrow \nu _{\tau }$ oscillations
measurable at accelerators. Cosmological bounds on other
rare processes, such as neutron-antineutron oscillations get also
relaxed by several orders of magnitude
compared with previous estimates.  }

 \Date{March 1992}
\noblackbox

\lref\man{N. S. Manton, Phys. Rev. D28 (1983) 2019\semi
F. R. Klinkhammer and N. S. Manton, Phys. Rev. D30 (1984) 2212.}
\lref\krs{V. A. Kuzmin, V. A. Rubakov and M. E. Shaposhnikov, Phys. Lett.
155B (1985) 36\semi
P. Arnold and L. McLerran, Phys. Rev. D36 (1987) 581; D37 (1988) 1020.}
\lref\ya{M. Fukugita and T. Yanagida, Phys. Rev. D42 (1990) 1285.}
\lref\hatu{J. A. Harvey and M. S. Turner, Phys. Rev. D42 (1990) 3344}
\lref\cdeo{B. A. Campbell, S. Davidson, J. Ellis and K. A. Olive,
Phys. Lett. 256B (1991) 457; CERN-TH.6208/91 (1992).}
\lref\fglp{W. Fischler, G.F. Giudice, R.G. Leigh and S. Paban, Phys.Lett.
258B (1991) 45.}
\lref\barr{S.M. Barr, R.S. Chivukula and E. Fahri, Phys.Lett. B241 (1991)
 387
    \semi  S. M. Barr, Phys. Rev. D44 (1991) 3062\semi
D. Kaplan, Phys. Rev. Lett.  68 (1992) 741.}
\lref\ird{L. E. Ib\'a\~nez and G. G. Ross, Nucl. Phys. B368 (1992) 3.}
\lref\thoo{G. 't Hooft, Phys. Rev. Lett. 37 (1976) 8;
 Phys. Rev. D14 (1976) 3422.}
\lref\shapdin{ For a review and references, see M.E. Shaposhnikov,
CERN-TH.6304/91 (1991) \semi P.B. Arnold, Argonne preprint ANEL-HEP-CP-90
-95 (1990).}
 \lref\nbb{A.E. Nelson and S.M. Barr, Phys.Lett. B246 (1991) 141.}
\lref\kss{S.Y. Khlebnikov and M.E. Shaposhnikov, Nucl.Phys. B308 (1988)
885.}
\lref\msw{For a review see S. Mikheyev and A. Smirnov,
Sov.Phys.Usp. 30 (1987) 759.}
\lref\mswr{A. Dar and S. Nussinov, Particle World  2 (1991) 117 \semi
L.E. Ib\'a\~nez, `Beyond the Standard Model (yet again)', CERN-TH.5982/91
, to appear in the Proceedings of the 1990 CERN School of Physics \semi
R.N. Mohapatra, preprint UMD-PP-91-188 (1991) \semi
S.A. Bludman, D.C. Kennedy and P.G. Langacker, preprint UPR 0443T (1991)
\semi  M. Fukugita and T. Yanagida, Mod.Phys.Lett. A6 (1991) 645.}
\lref\zwir{G. Costa and F. Zwirner, Rivista del Nuovo Cimento
9 (1986) 1.}
\lref\mohap{R.N. Mohapatra, Nucl. Instrum. Methods A 284 (1989) 1.}
\lref\tak{M. Takita et al., Phys.Rev. D34 (1986) 902.}

\newsec{Introduction}

 Non-perturbative Standard Model effects can play a fundamental role
 in particle interactions at high temperatures
\refs{\man,\krs}
         and  have to be
 taken into account in any description of the early history of the
 universe \shapdin  .
 The violation of $B+L$ (where $B$ and $L$ are the baryon and lepton
 numbers, respectively) from the anomaly structure of the weak interactions
 has led to  dramatic conclusions about the generation of the
 baryon-number
        asymmetry of the universe. In particular, if  there is a primordial
 generation of $B+L$  from a high-energy extension of the Standard
 Model (or as initial conditions), the electroweak processes generated by
 the anomaly  will erase any
 $B+L$ excess as long as the original $B-L=0$. Furthermore, if some effective
 interactions are generated by the high-energy theory which violate
 $B$ and/or $L$ and they are in equilibrium with the electroweak processes,
 then the whole baryon and lepton asymmetry disappears. Arranging that
 these different processes are out of equilibrium in terms of the expansion
 of the universe, puts strong constraints on the  couplings
 multiplying these interactions in the effective Lagrangian. In particular,
 this argument has been used in
 \refs{\ya,
         \hatu,   \nbb}    to  find
 very strong cosmological limits
 on neutrino masses and in \refs{
 \cdeo,  \fglp } to constrain  the
  non-renormalizable   $B$ and/or $L$-violating
           operators in supersymmetric models which are
 generically much stronger
 than any laboratory-based bounds.

 In this note we analyse these issues in the context
 of the supersymmetric standard model
 (SSM). We show  that all dimension-four (renormalizable) couplings
 in the SSM are invariant under two global ($R$) symmetries,
 contrary to the situation in the Standard Model, where there are
  no more global symmetries than $B$ and $L_i$. Similar to the situation
 of   $B+L$, these symmetries are anomalous. Convenient   combinations
 can be chosen such that one has $SU(2)$  mixed anomalies and the second
 has $SU(3)$ mixed anomalies. Actually they are further broken at the
 supersymmetry-breaking scale by the soft SUSY-breaking  and Higgsino
 mass terms; however, for sufficiently high temperatures
 (to be specified below),
       these are good global symmetries (up to anomalies)
                       which will induce
 non-perturbative operators through the anomaly, violating these
 symmetries together
 with $B+L$. By standard arguments  these couplings will not be
 exponentially suppressed at high temperatures \krs. Taking these
 interactions into account, we
 make an equilibrium thermodynamics analysis of the possible interactions
 of the SSM, and conclude  that the situation is drastically changed from
 previous studies. Contrary to the Standard Model case, primordial $B+L$
 is no longer erased, even for  $B-L=0$; instead the net excess
 of $B+L$ is partially converted into
 the generation of supersymmetric particles such as gauginos  and
Higgsinos (at least for $T\ge 10^7$ GeV).
In this sense, our mechanism can be seen
 as a  realization of a scenario
 proposed in \barr , for which extra global symmetries in extensions of
 the Standard Model could avoid erasing
 the baryon asymmetry of the universe.
There is an important difference though. In our case the
extra global symmetries are only good symmetries above a certain
temperature $T_{SS}\sim 10^{7}$ GeV, and hence provide only a
partial protection against the erasing of the primordial
baryon asymmetry. At temperatures below $T_{SS}$, the Standard
Model situation is recovered as far as baryon and lepton density
is concerned. Thus, in particular, a primordial $(B-L)$ generation
is required for a baryon asymmetry to remain at low energies.
              On the other hand we have found that the  different
situation for temperatures above $T_{SS}$ plays a crucial role
for determining the cosmological limits on non-renormalizable
 couplings violating baryon and/or  lepton numbers. In
particular we found that previous bounds on neutrino masses \ya\
            -      \fglp\
and neutron-antineutron oscillations \cdeo\ are relaxed for more than
five orders of magnitude, making the neutrino-masses limit
consistent with  an MSW+see-saw explanation of the number of solar
neutrinos.

 \newsec{Global Symmetries in the Supersymmetric Standard Model}

 We will consider the minimal supersymmetric extension of the Standard Model
 with gauge symmetry $SU(3)\times SU(2)\times U(1)$,
 three generations of quark ($Q, u_L^c, d_L^c$ ) and lepton ($L, E_L^c$)
 superfields, and two
 Higgs doublets  superfields ($H, \b H$). The gauge-invariant
 dimension-four operators that can appear in the superpotential are:
 \eqn\sup{\eqalign{W=h_u\, Q_Lu_L^c\b H + h_d\, Q_Ld_L^c H+h_l\, L_L E_L^c H\cr
 + h_B\, u_L^c d_L^c d_L^c +h_L\, Q_Ld_L^cL+h'_L\, L_LL_LE_L^c \ ,\cr}}
where generation and gauge indices have been suppressed.
Dimension-three operators like $\mu H\bar H $ and $\mu'LH$ can also
be included (although the latter can be set to zero by a field redefinition).
The first three terms in \sup\ provide masses to quarks and leptons,  whereas
the last three violate baryon or lepton number and their couplings
have to be restricted by phenomenological arguments such as the stabilty of the
proton.

It is well known that the Standard Model couplings are automatically
invariant under the four global symmetries $B, L_i,\ i=e, \mu , \tau$.
The anomaly of the combination $B+L$ ($L=\sum L_i$) is responsible for the
interesting physical  effects at high temperatures mentioned above.
In a similar way we can investigate the global symmetries of
the supersymmetric standard model (SSM)
        we just described. In fact this question was indirectly
discussed in the study of discrete gauge symmetries in ref.\ird\
with the conclusion that there are two global symmetries that
preserve all couplings in     \sup. In the notation of ref.\ird\ they
correspond to a regular symmetry $U(1)_A\equiv g_{PQ}$ and an
$R$ symmetry $U(1)_R\equiv I$. In table 1 we present the way they
act on the fermion components of the matter and gauge superfields
(we denote by a tilde the superpartner of a Standard Model field).
For $U(1)_R$ the bosonic component of the  superfield
has one unit more than the corresponding fermion.

Knowing how all the fields of the SSM transform under the symmetries,
it is straightforward to study the mixed anomalies with the
gauge symmetries. The $SU(N)-SU(N)-U(1)$ anomaly is given by
\eqn\anom{A=\sum_{R,i}{c_2(R)\mu_i q_i}\ \ ,}
where $c_2(R)$ is the quadratic Casimir operator in the representation $R$
of $SU(N)$ and the sum is also over the particles with charge $q_i$ and
 degeneracy
$\mu_i$ in the given representation. We are using standard conventions
for which $c_2(R)={1\over 2}$ in the fundamental representation and
$c_2(R)=N$ in the adjoint. For $U(1)_Y-U(1)_Y-U(1)$ anomalies
$A=\sum_i{y_i^2 q_i}$, where $y_i$  are the $Y$ charges of the particles.

 Using these expressions, we found that both symmetries have mixed anomalies
for any number of generations $\g$ and Higgs  pairs ($H,\b H$)  $\h$,
as shown in     table 1. It is convenient to define two independent
combinations $R_2\equiv I^{\g}\times g_{PQ}^{ 6-4\g}$ and $R_3\equiv
 I^{\g-\h}\times
g_{PQ}^{4+2\h-4\g}$ with the property that $R_N$ has only $SU(N)$
and $U(1)_Y$ mixed anomalies.
Their action on the fermionic components of the supermultiplets is
also shown in table 1.
Notice that the combination $B+L+R_2$ has no $SU(2)$ nor $SU(3)$
mixed anomalies for the physical case $N_g=3$. Thus, just like
$B-L$, the symmetry $B+L+R_2$ is absolutely conserved
(up to soft symmetry-breaking terms).

 Once supersymmetry is broken at a
 scale $M_{SS}$, these
global symmetries will also be explicitly broken by soft-breaking
terms such as    gaugino masses.  Also the dimension-3     operator in
the superpotential $\mu H\b H$ breaks $R_2$ and $R_3$ at a similar scale
($\mu\sim M_{SS}$). At sufficiently high temperatures (see below),
these symmetries
are exact up to anomalies, a situation analogous to that of
$B+L$ in the Standard Model.

Let us now restrict to the case where all $B$-  or $L$-violating
terms in \sup\ are forbidden. Therefore there will be a total
of six global symmetries $B, L_i, R_2$ and $R_3$, where $R_2$
and $B+L$ have mixed $SU(2)$ anomalies and $R_3$ has
mixed $SU(3)$ anomalies. In order to explore the physical
 consequences of these anomalies we have to find the effective
operators that they generate. For the Standard Model the
effective operator from the $(B+L)- SU(2)^2$ anomaly is \thoo\
\eqn\opu{O_1= (Q_L Q_L Q_L L_L)^{\g}}
where the three quarks have different colours to make an $SU(3)$
singlet and the power of $\g$ is actually a product over
generations. The structure of this operator can be understood as follows.
Each of the fields transforming non-trivially under $SU(2)$ enters
once, exactly in the way they contribute to the anomaly,  i.e.
proportional to the Casimir of the corresponding representation.
This corresponds to the standard counting of fermionic zero
modes in an instanton background.
The net effect of the anomaly is then to create this operator
out of the vacuum and then induce $B+L$ violation. Instantons can
create it at zero temperature but it is exponentially suppressed,
since it is a tunnelling effect, whereas at high temperatures there
is enough energy to go through the potential barrier and this
operator is believed to appear with an unsuppressed coefficient,
changing the net baryon number of the universe.

Following the same argument for the structure of the effective
operators generated by $SU(2)$ non-perturbative effects, one expects
in the SSM
an effective multifermion interaction:
\eqn\opd{O_2=(Q_LQ_LQ_LL_L)^{\g}(\t H\t{\b H})^{\h}\t W^4\ \ ,}
where now the $\h$ higgsinos ($\tilde H, \tilde{\bar H}$) and
the winos ($\tilde W$) transforming non-trivially under
$SU(2)$  also contribute to the anomaly and then to \opd ,
in a way proportional to the Casimir of the corresponding
representation. Notice that this operator is automatically
gauge, $B-L$,    $R_3$ and $B+L+R_2$ invariant, as it should. For the
$R_3-SU(3)^2$ anomaly the corresponding operator is
\eqn\opt{O_3=(Q_L Q_L u_L^c d_L^c)^{\g} \t g^6\ .}
Again \opt\ violates $R_3$ and preserves all the other
symmetries of the model. We therefore expect that \opd\
and \opt\ will play a role in the SSM at high temperatures
affecting the  baryon density of the universe.

One may wonder whether the existence of the above anomalous
global symmetries is just a property of the minimal version
of the supersymmetric standard model. We find that this   seems to be
quite a generic feature of SUSY extensions of the SM. For
example, one may consider a model with an extra singlet chiral
superfield $N$ with superpotential couplings $(NH\bar H)$ and
$N^3$. This has the advantage that the superpotential is
classically scale-invariant and one does not need to introduce
an ad hoc $\mu $ parameter. Again one can check that this model has a
$U(1)$ $R$-symmetry       which assigns the charges for chiral fermions
$Q_R(Q,U,D,L,E,\tilde H,{\tilde {\bar H}},\tilde N)=
(-1/6,-1,0,-5/6,2/3,-5/6,1/6,-1/3)$  and
$Q_R(\tilde g,\tilde W,\tilde B)=-1$ for gauginos. This symmetry has
both mixed $SU(2)$ and $SU(3)$ anomalies and the above arguments
      still apply.

 \newsec{Thermodynamic Equilibrium}

The question we will address now is, given a primordial excess
of $B$ and $L$, how    they evolve if the reactions induced by
\opd\ and \opt\ are in thermal equilibrium,  i.e.  they occur
faster than the expansion rate of the universe parametrized by
the Hubble constant.                    For this we need to
express the $B$ and $L$ number density in terms of chemical
potentials. Using the constraint that
all the SSM interactions be in thermal equilibrium, we
will find only a few  independent chemical potentials
on  which depend all the number densities.      Extra constraints
are obtained if the anomalous processes \opd\ and \opt\ are
also in equilibrium.

For ultrarelativistic particles, the equilibrium number density $\Delta n$
 (difference of particles and antiparticles) of a
particle species, depends on the
temperature $T$ and chemical potential $\mu$ of the respective
particles in the
following way:
\eqn\nude{{\Delta n\over s}=
{15\, g\over 4\pi^2 g_{*}}
              \big( {\mu\over T}\big )\cases{2\qquad{\rm bosons}\cr
              \noalign{\vskip2pt}
              1\qquad{\rm fermions}\,\cr}}
where $s={2\pi^2 g_{*}T^3 /    45}$ is the entropy density, $g$ is the
number of internal degrees of freedom and
$g_{*}$ is the total number of degrees of freedom ($\sim 200$  in the
supersymmetric case).
For simplicity of notation, we will name the chemical potential of
a given particle by the name of the corresponding particle.
Since we will work at scales much higher than $M_W$, we assume that
all the particles in the same $SU(3)\times SU(2)\times U(1)$
multiplet have the same chemical potential and that the corresponding
gauge fields have vanishing chemical potential. Because of
generation-mixing interactions, the quark  potentials will be taken
generation-independent. This will not be the case for the leptons
since lepton generation-mixing is not generically present  in the
absence of             right-handed neutrinos. Also, since
the interactions in \sup\ are assumed to exist for all the Higgses, they will
 have the same
chemical potential. Then, we have  to consider
$25=2\times 11 +3$ independent chemical potentials for the SM
particles (with two Higgses) plus their superpartners and the three
gauginos $\t W, \tilde g, \t B^0$.

The interactions in \sup\ imply
\eqn\thequ{\eqalign{u_L^c+Q_L+\b H & = 0\cr
                   d_L^c + Q_L + H&= 0 \cr
                   E_L^{ci}+ L_L^i +H &= 0\,}.}
Gaugino couplings of the SSM imply also:
\eqn\theqd{\eqalign{\t Q_L& = Q_L-\t g = Q_L - \t B^0\cr
                   \t L_L^i&=L_L^i-\t W=L_L^i-\t B^0\cr
                  \t H&=H+\t W=H+\t B^0\cr
             \t{\b H}&=\b H+\t W=\b H+\t B^0,}}
with similar relations for the right-handed quarks and leptons.
If all these processes are simultaneously in equilibrium, we
will have $\t B^0=\t W=\t g$.
We
are then left with the independent chemical potentials
$Q_L, L_L^i, H,\b H, \t g$.  In terms of these variables we
can express, using \nude, the baryon and lepton number densities
\eqn\baryon{\eqalign{B&={15\over 4\pi^2 g_{*} T}\{ \g\big( 2Q_L-u_L^c-d_L^c+
2(2\t Q_L-\t u_L^c-\t d_L^c)\big)\}\cr
&={{15\g}\over 4\pi^2 g_{*} T} \{ 12Q_L+3(H+\b H)-8\t g\}\ ,\cr
\noalign{\vskip2pt}
L&={15\over 4\pi^2 g_{*} T}\{\sum_{i}\big( 2L_L^i-E_L^{ci}+2(2\t L_L^i-
\t E_L^{ci})\big)\}\cr
&={{45\g}\over 4\pi^2 g_{*} T}\{ 3L_L+H-2\t g\}\ .}}
Here  we have defined $L_L\equiv {1\over{\g}}\sum_{i} L_L^i$.
We can also compute the total electric charge density $Q$ and
impose the constraint \hatu\ that it vanishes in a universe in  thermal
equilibrium. This implies
\eqn\charge{\eqalign{Q&={  {15}\over {4\pi^2 g_{*} T}}\{
2\g( Q_L-u_L^c+2(\t Q_L-\t u
_L^c))-\g(Q_L-d_L^c+2(\t Q_L-\t d_L^c))\cr
&\qquad\qquad-\sum_{i}(L_L^i -E_L^{ci}+2(\t L_L^i-\t E_L^{ci}))+2\h(\b
 H-H)+\h(\t{\b
H}-\t H)\}\cr
\noalign{\vskip2pt}
&={15\over 4\pi^2 g_{*} T}\{6\g(Q_L-L_L)+3(\h+2\g)(\b H - H)\}=0. }}
Of course, the same result is obtained if we imposse the
vanishing of the overall
  weak hypercharge instead of the charge.
Now we will use the condition that the $SU(2)$ anomalous couplings
obtained from \opd\ are also in equilibrium at high temperatures,
thus implying the relation
\eqn\aneq{3\g Q_L+\g L_L+\h(\t H+\t{\b H})+4\t W=0\  ;}
      if all the processes \thequ\ and \theqd\ are in equilibrium, this
reduces to:
\eqn\aneqp{3\g Q_L+\g L_L+\h (H+\b H)+(4+2\h)\t g=0.}
In a similar way, if the $SU(3)$ anomalous interactions \opt\ are in thermal
equilibrium, we will get the condition
\eqn\aneqd{2Q_L+u_L^c+d_L^c+2\t g=0 \, .}
  Again,  if the interactions \thequ\ and \theqd\
are in thermal equilibrium \aneqd\  reduces to
 \eqn\aneqt{  \g(H+\b H)-6\t g=0\, .}

The condition of vanishing electric charge, together with \aneqp , reduce the
number of independent variables to $Q_L, H$ and $\t g$. It is straightforward
to see that inserting    them into \baryon\    will imply a
non-vanishing $B+L$, even if $B-L=0$ as was the case for the SM; the
reason is    that we have more independent variables and the system of
homogeneous linear equations will then have a non-vanishing solution.
We believe however that, for consistency, the anomalous QCD couplings
also have to be in equilibrium, implying the further constraint \aneqd.
We will show explicitly that, even in this case, the same conclusion
holds. We have now only two independent variables $Q_L$ and $\t g$. Plugging
the conditions \charge, \aneq\ and \aneqd, we find the expression
for the baryon and lepton densities,
\eqn\barlep{\eqalign{B&={30\over 4\pi^2 g_{*} T}
\{6\g\,Q_L-(4\g-9)\,\t g\}\, ,\cr
\noalign{\vskip2pt}
L&=-{45\over 4\pi^2 g_{*} T}\{ {{\g(14\g+9\h)}\over{\h+2\g}}{Q_L}
+\Omega(\g,\h) \t g\} }\  ,}
where
\eqn\ratio{\Omega(\g,\h)={{2\g(2{\g}^2+6\g\h+3{\h}^2)+(14{\g}^2+39\g\h+18{N_h^2
) }}\over {\g
(\h+2\g)}}\, .}
We can easily see that even setting $B-L=0$ (which implies $Q_L=-{151/
 237}\t g$ for
$\g=3, \h=1$)  we will get a non-vanishing
$(B+L)\propto \t g\ne0$, indicating that
the baryon asymmetry does not disappear and the baryon excess partially
transforms into supersymmetric particles.
This is nothing but a reflection of the fact that, in the supersymmetric
case it is $B+L-R_2$ (and not just $B+L$) which is anomalous,
and $B+L+R_2$ is anomaly-free. Notice also that the fact that we are
left at the end with two independent chemical potentials
$Q_L$ and $\t g$ is a consecuence of the two  nonanomalous
symmetries $B-L$ and $B+L+R_2$ in agreement with the
fact that there is a chemical potential associated with each continuous
symmetry.
 In the case of an extended supersymmetric model with an extra
singlet $N$ as discussed in section 2, the same results as in    \barlep\
           hold. The only difference is the presence  of
equilibrium chemical potentials for the extra singlet
$N=-4\tilde g$, $\tilde N=2\tilde g$.

\newsec{Bounds on B/L-violating non-renormalizable operators}

In all the above we assumed unbroken supersymmetry
and no explicit Higgsino mass terms $\mu H\bar H$ in the
superpotential. These terms break explicitely the $U(1)_A$
                                               and $U(1)_R$
symmetries in table 1. Thus both $R_2$ and $R_3$ are
explicitely (but softly) broken by those two effects.
In particular, gaugino Majorana masses
$M_{\tilde g},M_{\tilde W},M_{\tilde B}$ and soft trilinear scalar
couplings proportional to the superpotential explicitely
break $R_2$ and $R_3$ (this is not the case of soft scalar
masses which preserve both), and the same is true for Higgsino
masses $\mu $. Let us take for simplicity all these soft terms
equal to a single symmetry-breaking parameter $M_{SS}\sim 10^2$
GeV. One would naively expect that for temperatures above $M_{SS}$
these symmetry-breaking effects would be negligible, in analogy
with what happens in spontaneously broken gauge symmetries such as
the electroweak. However the above symmetry-breaking is $not$ spontaneous
but explicit          and there is no reason to expect the
restauration of the symmetry above $M_{SS}$. Rather, the relevant
question is above which temperature             the effect
of the soft symmetry-breaking terms   falls out of thermal equilibrium.
This temperature, $T_{SS}$, turns out to be substantially higher than
$M_{SS}$.
     One can easily estimate this temperature,  above which,
      the symmetries  $R_{2,3}$ hold.
   To do that,   one    compares   the rate for these
symmetry-breaking effects, which may be estimated as
\eqn\rate{\Gamma _{SS}\ \simeq \ {{M_{SS}^2}\over T}}
with the expansion rate of the universe $\Gamma _H\simeq 30\times
T^2/M_{Planck}$. Then one finds that the $R_{2,3}$ and
supersymmetry-breaking
effects are outside thermal equilibrium for
\eqn\tss{T\ \geq \ T_{SS}\ \simeq \ {1\over {30^{1/3}}}\ M_{SS}^{2/3}
M_{Planck}^{1/3} \ \simeq \ 10^7\ \ GeV\ , }
where we have taken $M_{SS}\simeq 10^{2 }$ GeV for the numerical
evaluation.
Thus above this temperature the arguments given in the previous
section apply.
Below that temperature, the explicit gaugino and Higgsino masses
force  the gaugino chemical potentials to vanish and the results
of previous analyses are recovered.        In particular,
formulae \barlep \  coincide with those in     \kss\ and \hatu\ in
that limit (up to an overall factor).
 Thus if one wants      a baryon asymmetry
to remain, there must be, also in this case, a
$B-L$  asymmetry  created above the $T_{SS}$ scale.
The main difference between our results and  those in previous
analyses appears when there are additional $B/L$-violating
non-renormalizable couplings.

Let us      consider now how the present analysis is modified in the
presence of extra non-renormalizable interactions violating
$B$ and/or $L$ symmetries.
Some of those operators      are relevant for the generation of
phenomena such as neutrino masses or neutron-antineutron
oscillations, as we discuss below.  Any such operator
of dimension $D=4+n$ has the form
\eqn\nonren{O_n\ =\ {1\over {M^n}}\ (\Phi ...\Psi \Psi)\ ,}
 where $\Phi ,\Psi $ denote generic scalar and fermion fields of
the SSM                           and $M$ is some mass scale
characterizing the size of the interaction.
At temperatures below $M$, these interactions occur  at
a rate
\eqn\rated{\Gamma _n\ \sim \ {{T^{2n+1}}\over {M^{2n}}}}
and they are in thermal equilibrium at temperatures
\eqn\theqt{T\ \ge{  \big({30\, M^{2n}\over
{M_{Planck}}}\big )}^{1\over {2n-1}}\,.}
If these non-renormalizable interactions are in thermal
equilibrium, they lead to further equilibrium constraints.
                      In the absence of the extra
supersymmetric chemical potentials described above (i.e.
for $\t g =0$), the new equilibrium constraints
coming from $B$ and/or $L$  violating non-renormalizable
interactions will erase the primordial $B$- and $L$- asymmetries.
This was used in Ref. \cdeo\   to find very stringent
constraints  on the scales and couplings of those interactions.

In the supersymmetric case things change.
         For temperatures $T\ge T_{SS}$ the effect of
supersymmetric particles cannot be neglected,
i.e.       one has to consider in general a non-vanishing
gaugino chemical potential $\t g\ne 0 $. Thus, although
there is an extra constraint
for a given non-renormalizable $B/L$-violating coupling,
there is also a new free variable ($\tilde g$) in
the equilibrium equations. The net effect of this
is that, if there was  a non-vanishing primordial $B-L$,
it is not erased by the $B/L$ terms in thermal equilibrium
but it is only partially transformed into supersymmetric
particles. Thus the primordial $B-L$ density is safe as long
the $B/L$-violating terms are in thermal equilibrium $above$
the $T_{SS}$ scale. Below that temperature those interactions
have to be outside thermal equilibrium because the
protection provided by the supersymmetric partners
dissapears. Thus $M$ has to be accordingly constrained and one gets,
from    \theqt :
\eqn\eme{M\ \ge \ ({1\over {30}})^{   1 \over {2n}}\times
M_{Planck}^{   1 \over {2n}}\times
T_{SS}^{{2n-1}\over {2n}}\ ; }
    recalling    \tss ,  one finally gets in terms
of $M_{SS}$ :
\eqn\emed{M\ \ge \ ({1\over {30}})^{{n+1}\over {3n}}\times
M_{Planck}^{{n+1}\over {3n}}\times
M_{SS}^{{2n-1}\over {3n}}\,.}
Numerically,          for $M_{SS}=10^2$ GeV,  the following
lower bound on the mass parameter $M$ characteristic of
the non-renormalizable interaction  is found
\eqn\emet{M\ \ge \ 3^{-{{n+1}\over {3n}}}\times
10^{(7\ +\ 5/n)}\ GeV\ .}
This is to be compared with the equivalent expression obtained
without taking the supersymmetric effects into account  \cdeo\
\eqn\emec{M\ \ge \ 10^{(14\ +\ 2/n)}\ GeV\, .}
One can see that the effects of the supersymmetry protection
relaxes by several orders of magnitude the bounds obtained
for $M$. Let us now discuss some particular $B/L$ non-renormalizable
operators of some special phenomenological interest.

A particularly interesting case is that of the operator
\eqn\opn{O_{\nu }\ =\ {1\over M}\ (L_LL_L\bar H\bar H )_F \ ,}
which gives rise to Majorana neutrino masses upon
electroweak symmetry breaking of order
$m_{\nu }\simeq <\bar H>^2/M$. If $O_{\nu }$ is in thermal
equilibrium we will have the
extra chemical potential constraint
\eqn\const{L_L\ +\ \bar H \ =\ 0\ \ .}
Above the $T_{SS}$ temperature one finds then that
$\tilde g\ =\ -{{99}/     {59}}\ Q_L
$
and there is only one independent chemical potential, e.g.  $Q_L$.
Then one has $(B+L)\propto Q_L\propto (B-L)$, and as long as there
was (or is created) a $B-L$ excess, the $B$ asymmetry is not
erased. Below $T_{SS}$,  $\tilde g=0$ and the baryon
asymmetry disappears unless the $O_{\nu }$ interaction
gets outside thermal equilibrium. Imposing that condition
one gets
\eqn\neut{M\ \ge \ { 1\over {30^{2/3}}}\times M_{Planck}^{2/3}\times
M_{SS}^{1/3}\ \simeq \ 10^{12}\ GeV}
which in turn corresponds to a limit on (the heaviest) neutrino
mass
\eqn\neutd{m_{\nu }\ \le \ 10\ eV\ .}
This is to be compared with the much stronger limit \hatu\ -
\fglp,   $m_{\nu }\le 10^{-3}$ eV, obtained  ignoring the existence
of the additional global currents.
Physically these four orders of magnitude are very important,
since a $\tau $-neutrino mass of order $10$ eV would be compatible
with a MSW explanation \msw\ of solar neutrino data for see-saw-like
neutrino mass hierarchies \mswr\ . Furthermore it would allow for
terrestrial measurements of $\nu _{\mu }\leftrightarrow \nu _{\tau }$
oscillations.

              Concerning baryogenesis, this scheme is
compatible with a $B-L$-number generation at a temperature
$T\sim M\sim 10^{12}$ GeV, which is $not$ erased
in the region  $T_{SS}\le T \le M$, because it is protected
by supersymmetry, and    is $not$ erased
for $T\le T_{SS}$ either because the interactions $O_{\nu }$ are then
outside    thermal equilibrium.

A similar analysis may be done for the $\Delta B=2$ dim=7
SUSY-operator \zwir\
\eqn\onn{O_{n-\bar n}\ =\ {1\over {M^3}}\ (UDDUDD)_F\ ,}
where we denote $U\equiv u_L^c,D\equiv d_L^c$. This operator
gives rise (when conveniently `dressed') to
neutron-antineutron oscilations. If $O_{n-\bar n}$
is in thermal equilibrium one gets the chemical
potential constraint
\eqn\cons{2D\ +\ U\ =\ 2\tilde g}
which, when combined with the rest of the equilibrium
equations, yields $\tilde g=-11/5 Q_L$. Again, as long as an
initial $B-L$ excess is  present,  the $B$ asymmetry is not erased
above $T_{SS}$. Requiring that below $T_{SS}$ the interaction
$O_{n-\bar n}$ is outside thermal equilibrium leads to
\eqn\emes{M\ \ge \ ({1\over {30}})^{1/6}\times M_{Planck}^{4/9}
\times M_{SS}^{5/9}\ \simeq \ 10^9\ GeV\, .}

This lower bound is six orders of magnitude weaker
than the one computed \cdeo\  ignoring the extra global symmetries $R_2$,
 $R_3$.
After the `SUSY-dressing' of the operator $O_{n\leftrightarrow \bar n}$,
one gets a coefficient for the relevant 6-quark dimension-9
operator
$G_{n\leftrightarrow \bar n}\sim \alpha ^2  /(M^3M_{SS}^2        ) $.
The oscillation period is estimated to be $\tau _{n\leftrightarrow \bar n}
\simeq 1/\delta m$ with $\delta m\simeq G_{n\leftrightarrow \bar n}
(GeV)^{-5}\times (0.6\times 10^{-5})$ GeV  \refs{
\zwir, \mohap}.
Using the experimental
limit \tak\        $\tau _{n\leftrightarrow \bar n}\ge 1.2\times 10^8$ s.
,         one gets a bound $M\ge 10^8$ GeV.  This is quite close
to our cosmological limit and hence one cannot exclude the
observability of neutron-antineutron oscillations coming from
dimension-7 supersymmetric operators. This is to be contrasted
with the result obtained ignoring the above thermodynamic
analysis, which would rule out the observability of such effect
by 6 orders of magnitude. This shows us again  how the existence
of extra anomalous global currents above the $T_{SS}$ scale
leads to substantial modifications of previous phenomenological
constraints.

A similar analysis may be carried out for other
 $B$- and $L$-violating operators of dimension bigger than four.
Notice that the simultaneous presence of more than one
$B/L$-violating operator may be dangerous, since in this case
there will be in general more equations than free extra chemical
potential varieties;    this typically yields the equilibrium
condition $\tilde g=0$, and the whole analysis would be similar
to the non-supersymmetric case. The dim=5 operators $(QQQL)_F$ and
 $(UUDE)_F$ completely relax $\tilde g$ to zero,               but these
operators must    be          very much supressed anyway ($M\ge 10^{16}$
GeV),      since otherwise they would induce by themselves fast
proton decay.
                         The bound on the relevant dimensional
scale $M$ characterizing each operator may be obtained from    \emet .
One just gets $M\ge 10^{12},10^{10},10^{9}$ for dim=5,6,7
respectively, and from there one can easily estimate its
possible phenomenological relevance.

Notice also that the above considerations do not substantially
modify the situation concerning the bounds on
dimension-4    $ R$-parity-violating operators  \refs{
\cdeo,  \fglp}.
Indeed, for $renormalizable$
couplings, the strongest bounds are obtained for the
$B/L$-violating operators being in thermal equilibrium at
the critical $SU(2)$-breaking temperature $T_c$. Since
$T_c\ll T_{SS}$, at those temperatures the global $R_{2,3}$
symmetries are badly broken and the standard analysis
ignoring these effects applies. Thus in models with (non-negligible)
dimension-4     $R$-parity-violating couplings, there must exist
a low- energy baryogenesis (unless e.g. one of the lepton numbers
is conserved         ).

Let us briefly summarize the results obtained in this letter.
We have performed a thermodynamic  equilibrium analysis
concerning the baryon- and lepton- number densities in the
SSM.                           We argued that above a certain
temperature $T_{SS}\sim 10^7$ GeV, there are further global
(anomalous) symmetries beyond $B$ and $L$. At these temperatures
not only is $(B-L)$    anomaly-free but also $B+L+R_2$, and the
electroweak anomaly partially transforms a possible baryon excess
into supersymmetric particle (e.g. gaugino) density. This has
important consequences on the cosmological bounds on
$B/L$-violating non-renormalizable interactions recently
obtained by imposing  that a primordial baryon asymmetry
is not erased. Those bounds are relaxed by several orders
of magnitude and are then consistent with interesting measurable
effects in neutrino physics and other rare phenomena.
Supersymmetry is the ultimate cause of these modifications and
one can say that, in this sense, supersymmetry protects
(at least partially) the primordial baryon asymmetry.

\bigskip
\bigskip
\bigskip
 {\bf Acknowledgements}

We would like to thank                           C. Burgess, B. Gavela,
O. Pene, M. Quir\'os and M. Shaposhnikov for useful conversations.
One of the authors (FQ)
acknowledges the hospitality of the CERN Theory Division.

\listrefs
%

%
\vfuzz=2pt\baselineskip16pt\nopagenumbers
\def\Captionbox#1#2{$$\vbox{{}\break\vphantom{x}\hbox{#1}
\break\vphantom{x}\break\hbox{#2}\break\vir}$$}
\def\-{$$\phantom{-}$$}
\def\u#1{$$\vbox{{}\break\vphantom{x}\hbox{#1}\break\vir}$$}
\def\virt#1{\vrule height #1 width 0pt}
\def\vir{\hbox{\virt{3pt}}}
\def\$u{$\u}

%

%

$$\vbox{\offinterlineskip\halign{\strut#&
       \vrule#&\hfil$\,#\,$\hfil&
       \vrule#&\hfil$\,#\,$\hfil&
       \vrule#&\hfil$\,#\,$\hfil&
       \vrule#&\hfil$\,#\,$\hfil&
       \vrule#&\hfil$\,#\,$\hfil&
       \vrule#&\hfil$\,#\,$\hfil&
       \vrule#&\hfil$\,#\,$\hfil&
       \vrule#\cr
      \noalign{\hrule}
&&$\u{${\rm
Fermions}$}$&&\$u{$U(1)_A$}$&&\$u{$U(1)_R$}$&&\$u{$R_2$}$&&\$u{$R_3$
}$
&&$\u{$3B$}$&&\$u{$L$}$&\cr
\noalign{\hrule\vskip3pt\hrule}
&&$\u{$\t g, \t W,\t B$}$&&\u{\- 0}&&\u{\- 1}&&$\u{$\- \g$}$&&$\u{$\- \g-\h$}$
&&\u{\- 0}&&\u{\- 0}&\cr
\noalign{\hrule}
&&$\u{$Q_L$}$&&\u{\- 0}&&$\u{$-1$}$&&\$u{$-\g$}$&&\$u{$\- \h-\g$}$
&&\u{\- 1}&&\u{\- 0}&\cr
\noalign{\hrule}
&&$\u{$u_L^c$}$&&$\u{$-2$}$&&$\u{$-3$}$&&$\u{$\- 5\g-12$}$&&\$u{$5\g-\h-8$}$
&&$\u{$-1$}$&&\u{\- 0}&\cr
\noalign{\hrule}
&&\$u{$d_L^c$}$&&\u{\- 1}&&\u{\- 1}&&\$u{$6-3\g$}$
&&\$u{$\h-3\g+4$}$&&$\u{$-1$}$&&\u{\- 0}&\cr
\noalign{\hrule}
&&\$u{$L_L$}$&&$\u{$-1$}$&&$\u{$-1$}$&&\$u{$3\g-6$}$&&\$u{$3\g-\h-4$}$
&&\$u{$\- 0$}$&&\u{\- 1}&\cr
\noalign{\hrule}
&&\$u{$E_L^c$}$&&\u{\- 2}&&\u{\- 1}&&\$u{$12-7\g$}$&&
\$u{$8+3\h-7\g$}$&&\$u{$\- 0$}$&&$\u{$-1$}$&\cr
\noalign{\hrule}
&&\$u{$\t H$}$&&$\u{$-1$}$&&$\u{$-1$}$&&\$u{$3\g-6$}$&&\$u{$3\g-\h-4$}$
&&\u{\- 0}&&\u{\- 0}&\cr
\noalign{\hrule}
&&\$u{$\t{\b H}$}$&&\$u{$\- 2$}$&&\$u{$\- 3$}$&&\$u{$12-5\g$}$
&&\$u{$\h-5\g+8$}$&&\u{\- 0}&&\u{\- 0}&\cr
\noalign{\hrule\vskip3pt\hrule}
&&$\Captionbox{$SU(3)$}{${\rm\ anomaly}$}$&&$\u{$-{{\g}\over 2}$}$
&&$\u{$3-2\g$}$&&$\u{\- 0}$&&$\u{$
\h(\g-3)+\g$}$&&\u{\- 0}&&\u{\- 0}&\cr
\noalign{\hrule}
&&$\Captionbox{$SU(2)$}{${\rm\ anomaly}$}$&&$\u{$-{{\g-\h}\over
2}$}$&&$\u{$\h-2\g+2$}$&&$\u{$\h(3-\g)-\g$}$&&\u{\- 0}&&$\u{$
{{3\g}\over 2}$}$&&$\u{${{\g}\over 2}$}$&\cr
\noalign{\hrule}
&&$\Captionbox{$U(1)_Y$}{${\rm\ anomaly}$}$&&$\u{${{(6\h-5\g)}\over 6}$}$
&&$\u{$-{10\over 3}\g+\h$}$&&$\u{$\h(3-\g)-5\g$}$&&$\u{$2\h
-{{2 \g}\over 3}(5-\h)$}$&&$\u{$-{{3\g}\over 2}$}$&&$\u{$-{{\g}\over 2}$}
$&\cr\noalign{\hrule}
}}$$

\bigskip\nobreak
\centerline{Table 1 }
\bigskip
\noindent \vbox{\baselineskip 14pt \noindent
Charges of fermionic components of SSM superfields under the
global symmetries and mixed anomalies with gauge groups for
any number of generations $\g$ and Higgss pairs $\h$.}
\vskip8pt
\bye